\documentclass[a4paper]{article}
\usepackage{amsmath,amssymb,graphicx,caption,subcaption,float,cite}
\begin{document}
\title{Blockchain for steganography: advantages, new algorithms and open challenges}
\author{Omid Torki, Maede Ashouri-Talouki, Mojtaba Mahdavi \\
Department of Computer Engineering, University of Isfahan, Isfahan, Iran \\
 \texttt{\{o.torki, m.ashouri, m.mahdavi\}@eng.ui.ac.ir}}
\date{}
\maketitle
\begin{abstract}
Steganography is a solution for covert communication and blockchain is a p2p network for data transmission, so the benefits of blockchain can be used in steganography. In this paper, we discuss the advantages of blockchain in steganography, which include the ability to embed hidden data without manual change in the original data, as well as the readiness of the blockchain platform for data transmission and storage, which eliminates the need for the Steganographer to design and implement a new platform for data transmission and storage. We have proposed two algorithms for steganography in blockchain, the first one is a high-capacity algorithm for the key and the steganography algorithm exchange and switching, and the second one is a medium-capacity algorithm for embedding hidden data. Also, by reviewing the previous three steganography schemes in blockchain, we have examined their drawback and have showed that none of them are practical schemes for steganography in blockchain. Then, we have explained the challenges of steganography in blockchain from the steganographers and steganalyzers point of view.
\end{abstract}

\textit{\textbf{Keywords---}} Steganography, Blockchain, Bitcoin, Transaction Address, Hierarchical Deterministic Wallets
\section{Introduction}\label{sec.introduction}
In addition to cryptography, which is aimed to hide the exchanged information, steganography is created to hide the principle of existence of communication between two persons\cite{morkel2005overview}. In steganography, if the adversary even doubts the existence of a communications (while he has not even been able to prove it), the steganographer has failed. In some steganography schemes, for improve the security, the message is first encrypted and then embedded, which is contrary to the nature of steganography. At the other side, steganalysis is used to discover the existence of a communication. Any media can be used for steganography, but media with a high degree of redundancy are more suitable\cite{morkel2005overview}. For this reason, photo, audio and video are often used for steganography\cite{imagesteganography,audiosteganography,videosteganography}.

Blockchain is a p2p network firstly used for digital currency\cite{nakamoto2008bitcoin}. Due to its unique features, researchers in various fields have taken advantage of the Blockchain\cite{blockchainusage1,blockchainusage2,blockchainusage3,blockchainusage4,blockchainusage5,blockchainusage6}. Bitcoin is the first and most widely used digital currency. In bitcoin, the distributed consensus achive between miners ensures that the information sent to the blockchain remains unchanged and permanent in the blockchain. In bitcoin, the sender signs the transaction and sends it to the blockchain. There are also different payment models in bitcoin, the three most popular of which are pay to public key(p2pk), pay to public key hash(p2pkh) and pay to script hash(p2sh). In p2pk, the sender deposits the money into the receiver's public key, and the receiver can receive it by signing his public key and sending it to the blockchain. In p2pkh, the sender deposits the money into the receiver's public key hash, and the receiver can receive the money by sending the public key and signing it to the blockchain. In p2sh, the sender and the receiver agree on a script and send its hash to the blockchain, and when they want to execute this script, they execute it by sending it to blockchain. Each bitcoin transaction contains one or more input addresses and one or more output addresses, meaning that a bitcoin amount is collected from one or more input addresses in a transaction and credited to one or more output addresses.

In this paper, we introduce the advantages of blockchain for embedded hidden data(steganography) and propose two algorithms for this purpose, which include high-capacity algorithm and medium-capacity algorithm for data embedding. The unique feature of these algorithms is that they do not manually change the original data during the embedding process, which makes the steganalysis very difficult. In addition, by analyzing previous steganography schemes in blockchain, we will show that there are no practical schemes for this purpose.

The organization of this paper is as follows: Section \ref{sec.Blockchain as a platform for steganography} introduce the benefits of blockchain in steganography. In Section \ref{sec.Analysis of previous steganography schemes in blockchain} we review the previous schemes for steganography in blockchain and analyze them. In Section \ref{sec.Hierarchical Deterministic Wallets} we review the HDW algorithm, that is a building block for our proposed algorithms. In Section \ref{sec.SSB: new Secure algorithms for Steganography in Blockchain} we propose our two algorithms for steganography in blockchain and in Section \ref{sec.evaluation} we evaluate the proposed algorithms. Section \ref{sec.Open Challenges} describe the open challenges of steganography in blockchain and finally Section \ref{sec.conclusion} concludes the paper.
\section{Blockchain as a platform for steganography}\label{sec.Blockchain as a platform for steganography}
Since blockchain is a p2p network available from all over the world, it will be a great platform for steganography. Specifically, the blockchain has two major advantages over its previous platform for steganography such as image and video. The first advantage is that for steganography in an image or video, we need to change parts of it to get the image or video we want. However in blockchain, it is possible to obtain transactions that, for example, the recipient's address contains the information we want (by repeating and creating different addresses to reach the desired address) or permutation addresses that have no specific order, we have embedded our data while we have not made any changes. The second advantage is that in the old methods of steganography, it is necessary to send an image and a video or a place to store it, and the repeated sending of these files is self-doubtful, while in the blockchain, due to its inherent nature, sending and receiving transactions are common and the sender and receiver of data do not need to design and implement a platform to send and store their data and simply use the ready-made blockchain platform.
\section{Review and Analysis of previous steganography schemes in blockchain}\label{sec.Analysis of previous steganography schemes in blockchain}
 In this section, we review and analyze the proposed schemes of Xu et al.\cite{xu}, Zhang et al.\cite{zhang} and Partala et al.\cite{partala}.

In the Partala et al. scheme\cite{partala}, the sender puts one bit of the data in the least significant bit(lsb) of the transaction address(the output address) and sends it to blockchain. This process repeats until embedding the all bits of the data. In order to  ensure correct retrieval of information at the receiver side, the sender should be wait for each transaction to be placed in the blockchain and then he can send the next bit to the blockchain through the next transaction. So, this scheme needs one hour and twenty minutes to send a byte of information through the bitcoin network. The receiver must also have the sender's addresses in order to detect the entry of information into the blockchain. Therefore, the sender need to use a small number of addresses that receiver is aware of it, which makes it easier to track.

In the zhang et al. scheme \cite{zhang}, the sender ﬁrstly encrypts the data and then	encodes	it by the base-58 encoding	system in order	to	have similar encoding as bitcoin addresses. Then, he selects a number of its data characters and generates an address that contains these characters	using the vanitygen	software. He repeats this
process until he obtains a number of addresses that	contain hidden data. Then, using an indexing algorithm, it generates an index for	the correct	retrieval of information at	the receiver side and encrypts and places it	in the op-return command. The op-return is a command that prevents the subsequent data to be processed by the miners, similar to the comments in the programming languages. The ﬁrst challenge in this scheme is that	it	does
not	explicitly	state how the receiver	detects the presence of new	information in the blockchain. Most	likely, the receiver should	check and decrypt all transactions that	have an	op-return command and see if there is new data. However, the second and	most important challenge is	the	use	of	op-return command; op-return is	a low-usage	command	and	putting the	encrypted data in it is highly questionable. Also, since the adversary knows the algorithm,	the steganalysis is not very complicated.

In the Xu et al. scheme\cite{xu}, the sender must be a miner. In this scheme, the miner, by using a key, firstly selects a number of transactions of a block that he intends to create and publish in blockchain, afterward he uses the permutation of these transactions to embed the hidden data in the block. Nowadays, a large number of miners are working together to create a pool, such as BTC, AntPool and single miner has no power in the blockchain and is unable to produce new blocks. On the other hand, only manager of a pool determines the order of a block transactions. Due to the small number of pools, only a few pool managers in the world can use this method for steganography, however, since pool managers have a lot of capital and power, they do not need to use steganography, so this scheme is completely impractical.
\section{Hierarchical Deterministic Wallets(HDW)}\label{sec.Hierarchical Deterministic Wallets}
\begin{figure}
  \centering
  \includegraphics[scale=0.8]{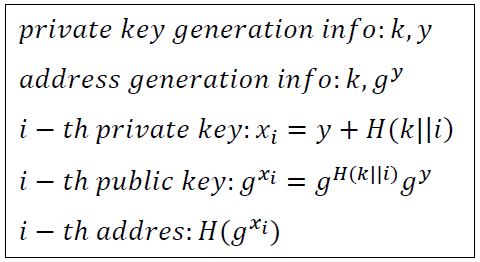}
  \caption{HDW algorithm}\label{fig.HDW algorithm}
\end{figure}

Hierarchical Deterministic Wallets(HDW) is a method for creation of a public key, a private key and a transaction address in bitcoin and other digital currency. In HDW, instead of having a private key and an address, we have a piece of private key generation info and address generation info that can be used to generate unlimited number of private keys and addresses\cite{bitcoinbok}. The advantage of this method is that, it is possible to generate unlimited number of private keys, public keys and addresses, which, although generated for us on a regular basis and using an algorithm, are completely random and unrelated to those who are not aware of the key. This algorithm works as follows\cite{bitcoinbok}: the private key generation info are $k,y$ and the address generation info are $k,g^y$. The i-th private key is obtained as $x_i=y+H(k||i)$ and the i-th public key is obtained as $g^{x_i}=g^{H(k||i)}g^y$, where $H$ is a hash function and $"||"$ means concatenation. For the p2pk payment model, the i-th public key is sufficient, but for the p2pkh payment model, the i-th address is the hash of i-th public key. The algorithm is shown in Figure \ref{fig.HDW algorithm}.

\section{SSB: new Secure algorithms for Steganography in Blockchain}\label{sec.SSB: new Secure algorithms for Steganography in Blockchain}
In this section, we introduce two algorithms for steganography in blockchain. The first algorithm is a high-capacity embedding algorithm that is used for key agreement, key exchanging, steganography algorithm agreement and so on, while the second algorithm is medium-capacity embedding algorithm for hidden data embedding and transmission.
\subsection{high-capacity embedding algorithm}\label{sec.high-capacity embedding algorithm}
This algorithm has high embedding capacity and is used for key agreement, key exchange, steganography algorithm agreement and similar cases that require more embedding capacity. In this algorithm, the user will lose some bitcoins. In this algorithm, the sender encrypts his message with a symmetric encryption algorithm such as AES and sends it to the blockchain through the p2pkh or p2sh payment model(of course, before sending it to the blockchain, sender should be use a random padding in order to match the bit length of the encrypted message with the output bit length of the hash that used in bitcoin). Since the output of the symmetric encryption and the hash function are both random and indistinguishable from each other, it is impossible for the adversary and people who do not know the key to recognize the transaction that is not the output of the hash function. The transaction input address must be created with the HDW algorithm to notify the receiver that a new message will be sent to the blockchain.
Clearly, the sender loses the bitcoin that entered in the transaction in this algorithm and cannot use it in future transactions.
\subsection{medium-capacity embedding algorithm}\label{sec.medium-capacity embedding algorithm}
This algorithm has less capacity for embedding and is used for hidden data embedding. The details of this algorithm are as follows: the sender first decides how many output addresses he intends to insert in the transaction
and how many bits he intends to embed in each address, then selects the desired bits of each address to embed the hidden data.  Afterward, using the HDW algorithm and by testing different values for $i$, the sender obtains the desired address that contains the embedded bits. To obtain the address that contains $m$ bits of desired data, we need an average of $2^ m$ effort(run HDW algorithm). Once all of the addresses have been created, more bits can be embedded by permuting these addresses. For this purpose, $n$ addresses have $n!$ permutations, that we must assign a number from zero to $n! -1$ to each permutation. To achieve this goal, it is enough to have a solution to compare between two permutations of addresses. We use the same method as\cite{xu} for this purpose, so that if we have two permutations $(a_1, a_2, ..., a_n)$ and $(a'_1, a'_2, ..., a'_n)$, we say $(a_1, a_2, ..., a_n)$ is greater than $(a'_1, a'_2, ..., a'_n)$, if we have $i\in [1,n]$ that:\\
\[
     \begin{cases}
       \text{$a_j=a'_j$} &\quad\text{$j < i$}\\
       \text{$a_j > a'_j$} &\quad\text{$j=i$} \\
     \end{cases}
\]\\
Suppose that each transaction contains $n$ output addresses and embeds $m$ bits in each address. Using this algorithm, the number of embedded bits in the addresses is equal to $nm$ bits and the number of embedded bits arising from the permutation of the addresses is $\log_2 n!$, and the capacity of this algorithm is equal to $nm + \log_2 n!$ per transaction. In this algorithm, the bitcoin that entered in the transaction can be used in subsequent transactions and the user does not lose money. It is worth noting that at the beginning of the data transmission between sender and receiver, the value of the variable $i$ in the HDW algorithm is equal to one and increased in order. The receiver can run the HDW algorithm and testing different values of $i$, starting with one, to notify the presence of new hidden data. Also, at appropriate intervals, the sender and receiver can change the keys through the high-capacity algorithm.
\section{evaluation}\label{sec.evaluation}
In\cite{morkel2005overview,petitcolas1999information}, four criteria for evaluating steganographic algorithms are introduced and we evaluate our algorithm with these criteria.\\
1) Visibility: This means that the information that contains hidden data is indistinguishable from the information without hidden data by the eye. In our scheme, this feature is maintained due to the use of HDW algorithm. In addition, in our algorithm, none of the address bits are changed manually, instead by repeating the HDW algorithm, we reach the desired address, so the address containing hidden data has no different from the address without hidden data, even with statistical analysis it is indistinguishable.\\
2) Robustness: This means that the embedded information is not lost due to an inadvertent or intentional changes and can be retrieved. In the proposed algorithms, due to the distributed consensus protocol and the signature on the transactions, the data remains permanently unchanged in the blockchain.\\
3) Security: This means that no one is aware of the existence of hidden data. As stated in the proposed algorithms, the data is not embedded manually and the address containing the hidden data is indistinguishable from the address without hidden data.\\
4) Capacity: The maximum data that can be embedded. As stated in Section \ref{sec.SSB: new Secure algorithms for Steganography in Blockchain}, the capacity of the high-capacity algorithm is equal to the output bit length of the hash function in Bitcoin, and the capacity of the medium-capacity algorithm is equal to $nm+\log_2 n!$ bits.

\begin{figure}
  \centering
  \includegraphics[scale=0.5]{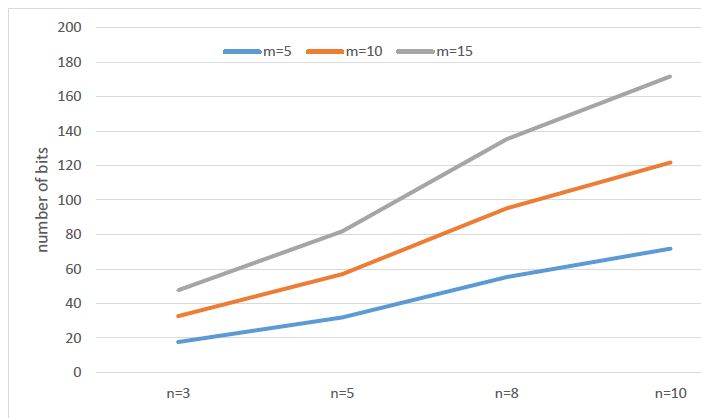}
  \caption{capacity of medium-capacity algorithm}\label{fig.capacity of medium-capacity algorithm}
\end{figure}

Based on selecting 30 random transactions from 10 randomly selected blocks from\cite{btc}, the number of output addresses of each transaction has an average of 3.45 and a standard deviation of 1.2. However, 5 output addresses are common in each transaction and in the review of 30 transactions, 12 transactions had 5 output addresses. There are also transactions in the blockchain with even more than 30 output addresses and the number of output addresses of transactions can be different at different times, one of the reasons is the difference in the transaction fee at different times. The time of embedding the hidden data in the output address of the transactions by using the HDW algorithm (obtaining the desired address) in terms of the number of bits that we intend to embed $m$ is shown in Table \ref{table.time of embedding hidden data using HDW algorithm}. Implementation performed on an Intel (R) Core (TM) i5-6200U processor with 8GB memory, running Windows 7 and Python programming language. The number of bits that can be embedded in a transaction(capacity of medium-capacity algorithm) for different values of the number of bits embedded in each transaction,$m$, and the number of output addresses of each transaction, $n$, is shown in Figure \ref{fig.capacity of medium-capacity algorithm}. Therefore, for security reasons, 81.9 bits can be embedded in a transaction with 5 output addresses and 15 embed bits per address.
\begin{table}
\caption{time of embedding hidden data using HDW}\label{table.time of embedding hidden data using HDW algorithm}
  \centering
  \begin{tabular}{c| c c c}
    \hline
    number of bits(m) & 5 & 10 & 15 \\ \hline
    time(s) & 0.0002 & 0.05 & 11.1 \\
    \hline
  \end{tabular}
\end{table}
\section{Open Challenges}\label{sec.Open Challenges}
As mentioned earlier, blockchain is a very suitable platform with unique features for steganography. However, this field is just beginning and there are many challenges that can be addressed in a variety of ways. More specifically, there are two major challenges, one for steganographers and one for steganalyzers:\\
-Finding blockchain features (in all digital currencies, not just bitcoin) that can embed data and provide high-capacity algorithms is an open challenge for steganographers, especially embedding algorithms, as stated earlier, should be able to embed the original data without manual change and to repeat until the desired data is reached.\\
-Finding methods to discover steganography in blockchain will be an open challenge for steganalyzers, especially since the new steganography algorithms embed the data (like the algorithm presented in this paper), without manually changing the original data and only by repeating the process until reaching  the data they want.
\section{Conclusion}\label{sec.conclusion}
In this paper we describe the advantages that blockchain can have for steganography and prove that blockchain can serve as a very good steganography platform, especially as steganography can be carried out in blockchain without altering the original data and only can be done by repeating until that the data is embedded and the blockchain itself is a ready-made platform for data transmission and storage, and Steganographer does not need to design a new platform for data transmission and storage. In the following, by examining the previous three schemes for steganography in blockchain, we showed that each of them has problems and none of them are practical. We propose two algorithms for steganography in blockchain, one is a high-capacity algorithm for exchanging keys and algorithms, and the other is medium-capacity algorithm for embedding hidden data. Finally, the challenges posed to steganography in blockchain for steganographers and steganalyzes are raised.

\bibliographystyle{ieeetr}
\bibliography{Torki}
\end{document}